# Carbon solubility and superconductivity in $MgB_2$

A. Bharathi[*], S.Jemima Balaselvi, S. Kalavathi, G. L. N. Reddy, V. Sankara Sastry, Y.Hariharan and T.S. Radhakrishnan

Materials Science Division, Indira Gandhi Centre for Atomic Research, Kalpakkam, 603102

Abstract

Successful replacement of B by C in the series $MgB_{2-x}C_x$ for values of x upto 0.3 are reported. Resistivity and ac susceptibility measurements have been carried out in the samples. Solubility of carbon, inferred from the observed change in the lattice parameter with carbon content indicates that carbon substitutes upto x=0.30 into the $MgB_2$ lattice. The superconducting transition temperature, $T_C$ measured both by zero resistivity and the onset of the diamagnetic signal shows a systematic decrease with increase in carbon content upto x=0.30, beyond which the volume fraction decreases drastically. The temperature dependence of resistivity in the normal state fits to the Bloch-Gruneisen formula for all the carbon compositions studied. The Debye temperatures, $\theta_D$, extracted from the fit is seen to decrease with carbon content from 900K to 525K, whereas the electron-phonon interaction parameter, $\lambda$, obtained from the McMillan equation using the measured $T_C$ and $\theta_D$, is seen to increase monotonically from 0.8 in $MgB_2$ to 0.9 in the x=0.50 sample. The ratio of the resistivities between 300K and 40K versus $T_C$ is seen to follow the Testardi correlation for the C substituted samples. The decrease in $T_C$ is argued to mainly arise due to large decrease in $\theta_D$ with C concentration and a decrease in the hole density of states at $N(E_F)$.

## INTRODUCTION

The recent discovery of superconductivity at 39K in $MgB_2$ has initiated a flurry of activity aimed at understanding the origin of the large $T_c$[1]. Several experimental studies, notably the isotope effect[2] and the pressure dependence[3] of $T_c$ point to the fact that superconductivity in this system is largely governed by the conventional electron-phonon mechanism and that the high $T_c$ is a consequence of the large hole density of states arising from the sigma bands of Boron and high phonon frequency arising on account of the low mass[4,5,6]. There have been several theoretical predictions of an increase in $T_c$ by chemical substitutions at the Mg and B sublattice [6,7]. There have been experimental reports of substituting Mg by Al[8,9], Li[10,11], Si[10], 3d transition metals[12,13,14] and recently 4d transition metals[15]. Almost all substitutions have led to a decrease in $T_c$, with an exception of Zn[14], which shows no change even for 30% substitution. Our study has shown that ~5% substitution of Nb[15] at the Mg site, in $MgB_2$, results in a small albeit definite increase in $T_c$ by ~0.5K. Of all the substitution studies carried out on the Mg sublattice, that of Al has been the most detailed. Slusky et. al[8] show in their study on $Mg_{1-x}Al_xB_2$ that Al substitutes upto x=0.1, beyond which a two phase region is observed consisting of the same $AlB_2$ structure differing only by the c-parameter. Their primary conclusion is that the large $T_c$ in this binary intermetallic arises because of the proximity of $MgB_2$ to a structural phase transition. Pressure studies[16] carried out upto 42 GPa do not give any evidence for a phase transition. Li et. al[9] show that in $Mg_{1-x}Al_xB_2$ there are two structural phase transitions at x=0.17 and x=0.75 and that there is an ordered compound at x=0.5, consisting of alternate layers of Mg and Al, corresponding to the composition $MgAlB_4$, that has a $T_c$ of 12K. Thermopower studies[17] on the $Mg_{1-x}Al_xB_2$ series reveal that the decrease in $T_c$ is related to the decrease in the hole density of states, the magnitude of which is in accordance with band structure calculations. Very recent theoretical investigations[18] have shown that the decrease in $T_c$ in the $Mg_{1-x}Al_xB_2$ system can be understood solely based on the decrease in the hole density of states due to electron doping on account of Al substitution, the occurrence of the ordered phase has been reproduced by the calculations.

In $MgB_2$, the substitution of the boron sublattice with C ($MgB_{2-x}C_x$), has been carried out by several authors[19,20,21,22], but the results are at variance with each other. While Takenobu et. al[19] show that $T_c$ decreases with C substitution upto the solubility limit of x=0.10, beyond which superconductivity disappears, Paranthaman et. al[20] claim that $T_c$ remains unchanged with C substitution, and that the volume fraction of the superconducting phase decreases with carbon substitution. They conclude that C is insoluble in $MgB_2$, and that carbon precipitation at the grain boundaries is responsible for the decrease in the superconducting volume fraction. The results of Ahn et.al[21] are similar to that of Takenobu et.al[19] but the $T_c$ depression due to carbon substitution is much smaller than that observed by the former for x=0.1. The origin of the difference in the C substitution behaviours can be attributed to arise on account of the difference in the starting materials used in the synthesis. It appears that the solubility is larger if the starting material is amorphous carbon or soot[19,21]. The initial interest in our study was motivated by recent theoretical calculations that predict the possibility of an increase in $T_c$ by chemical substitution on the Mg sublattice by a monovalent cation viz., Cu, to increase the hole concentration, along with a C substitution in the B sublattice to increase the stiffness of the Boron layer[6]. Since the carbon substitution results were controversial we first re-examine the C solubility in $MgB_2$ by preparing samples of good quality under high Ar pressure and study its effect on $T_c$ both by resistivity and susceptibility. Further we analyse the resistivity data in the normal state, using the Bloch Gruneisen formula and show that the Testardi correlation[23] is verified in the carbon series, which was hitherto not feasible[21].

**EXPERIMENTAL**

For the synthesis of materials we employed a novel method of sealing the initial mixture under a high pressure of Ar (Hinks et. al)[24], the procedure employed by us is described in the following. Stoichiometric quantities of Mg (99.9%), amorphous Boron (99%) and carbon soot (99%), obtained as a by-product in fullerene synthesis, were ground, pelletised and loaded into a Ta crucible which was subsequently placed into a thick walled SS tube closed at the bottom. After a few initial evacuations and Ar flushings, the SS tube is filled with high purity (99.998%) Ar gas at a pressure of 35 bar. With the help of a high pressure valve this pressure is locked inside the tube. The SS tube assembly is fitted via a Wilson seal to an outer quartz tube, which is evacuated to $10^{-5}$torr. The entire assembly of quartz tube and SS tube containing the sample is inserted into a vertical furnace kept at $900^{o}$C. The heat treatment is carried out for 1 hour and 30 minutes, during which the Ar pressure is seen to rise to 50 bar. After the heat treatment the furnace is switched off, and the cold sample is removed from the SS tube and weighed. No Ar gas leak was observed during this heat treatment. The weight loss of the sample after the heat treatment was negligible, recorded to be routinely < 1%. Diamagnetic signal, from 25 mg lots of the powdered samples were recorded by the ac susceptibility technique , at a measuring frequency of 941Hz. The resistivity measurements were carried out in the standard four probe geometry, on chunks of the sample which were dense enough to take electrical contacts with silver paste. Our efforts to pelletize and sinter the samples resulted in them becoming insulating. The temperature variation in the 4.2K to 300K temperature range was achieved in a dipstick arrangement and data collected by a PC using the IEEE interface card. XRD measurements were carried out in the Bragg-Brentano geometry in the $9\text{-}70^{o}$, $2\theta$ range at a step size of $0.1^{o}$, in a STOE diffractometer, using Cu $K_{\alpha}$ radiation.

**RESULTS**

Fig.1a shows the XRD patterns of $MgB_{2-x}C_x$ obtained from the heat treatment under 50 bar of Ar. It can be seen that all the peaks can be indexed to $MgB_2$ lines corresponding to the hexagonal, P6/mmm structure for concentration upto x=0.30. Analysis of the XRD pattern in $MgB_2$ reveals that the lattice parameters are a=3.0846$A^{o}$ and c=3.5224$A^{o}$, which are in good agreement with the reported values in literature[1,2]. The MgO fraction is estimated to be ~1% using the PCW program. XRD showed the absence of $MgB_4$ in all the samples, which is an indication of the lack of evaporation of Mg during the heat treatment on account of the high pressure in Ar gas environment. The XRD results reveal (see Fig.1b) that the (100) lines show a systematic shift with carbon substitution upto x=0.30. Beyond this concentration there is no shift in the peak and it does not show splitting even at a carbon fraction of x=0.5, in contrast to the results in $Mg_{1-x}Al_xB_2$, indicating the absence of any phase separation. In Fig.2 is shown the variation of the lattice parameters and the cell volume with carbon substitution, as obtained from an analysis of the line positions using the STADIP program. The c-axis shows a small increase from 3.523 $A^{o}$ in pristine $MgB_2$ to 3.527 $A^{o}$ at x=0.3, beyond which it decreases to 3.519$A^{o}$ (cf. Fig. 2b). Correspondingly, the a-axis (cf. Fig. 2a) shows a larger decrease with increase in C fraction viz., from 3.084 $A^{o}$ to 3.0505$A^{o}$ for x=0.3, beyond which, it remains constant. The a-axis variation is similar to the earlier report of Takenobu et. al[19] for x<0.08, but the continued decrease beyond this composition seen in Fig.2a, clearly suggests the enhanced solubility of C in our samples and the absence of phase separation. The cell volume variation (cf. Fig. 2c) with carbon fraction follows a variation very similar to that obtained for the a-axis.

In Fig.3 is shown the variation of the diamagnetic signal for the various carbon substitutions in the 4.2K to 50K temperature range. It can be seen that superconducting transition is sharp in $MgB_2$, with a width of ~1.5 K and that the $T_c$ onset decreases with C substitution. The magnitude of the diamagnetic signal at the lowest temperature measured, is independent of the carbon concentration for x<0.15. For larger carbon concentrations, the superconducting transitions broaden, and the saturating signal at low temperatures decreases, but not drastically. These results suggest that bulk superconductivity persists even for carbon substitutions of x=0.3. Identifying $T_{c,}$ to be the temperature at which the signal value in each sample attains the 90% of its signal value, the $T_c$ is evaluated for all C substitutions and is shown as a function of carbon fraction Fig.4. It is clear from the figure that the $T_C$ decreases with C fraction linearly for x=0.3, beyond which it flattens off. The variation of resistivity with temperature in 4.2K to 300K region is plotted for four samples studied in Fig.5. It is clear from the figure that the resistive transitions are much sharper than those seen by diamagnetic susceptibility, having a transition width of ~0.3 K in pristine $MgB_2$. The RR defined as the ratio of resistivity at 300K to that at 40K is large, ~6.5, in pristine $MgB_2$. The $T_c$ obtained from resistivity measurements as the temperature of the downset of the transition is 39.4K in pristine $MgB_2$. The $T_c$ obtained similarly for all carbon compositions studied is shown along with that obtained from diamagnetic susceptibility in Fig.4. It can be seen from the figure that the $T_c$ obtained by both the experiments are in agreement. The narrow transition width, the high $T_c$ onset and the large RR, in $MgB_2$ clearly indicate the good quality of the samples. With increase in carbon content the RR shows a systematic decrease. It has been proposed that like in conventional Nb based superconductors the Testardi correlation is valid in $MgB_2$, and is thought to be yet another evidence for the fact that the electron-phonon mechanism is operative in determining superconductivity in $MgB_2$[23]. The variation of RR versus $T_C/T_m$, where $T_m$ is the maximum $T_c$ observed in $MgB_2$, in the four carbon substituted samples studied by resistivity have been plotted with similar data in $MgB_2$[25] in Fig.6. It can be seen from the figure that the carbon substituted data also conform to the Testardi correlation, implying that the electron-phonon interaction parameter is primarily responsible for superconductivity in this system.

The temperature dependence of resistivity in the normal state has been fitted to the expression

$$R(T) = A + BT^2 + KR_{ph}(T)$$

Where

$$R_{ph}(T) = (m-1)\theta_D (T/\theta_D)^m \int_0^{\theta_D} dz \frac{Z^m}{(1-e^{-z})(e^z-1)}$$

A is the impurity scattering contribution to resistivity. B is the coefficient describing the magnitude of the electron-electron scattering, m=5 and $\theta_D$ is the Debye temperature and K is a constant. Fitting the resistivity data to this expression results in excellent fits (see inset Fig.5) and the extraction of $\theta_D$, A, B and K. In pristine $MgB_2$, A is $4.4\times10^{-05}\Omega$, the coefficient B turns out to be $2.3\times10^{-10}\Omega T^{-2}$ and K is $3.9\times10^{-3}\Omega$, implying that electron-electron scattering is negligible in determining transport in this system. Using $\theta_D$, obtained from the resistivity data and the experimentally obtained $T_c$, in the McMillan equation

$$T_c = \frac{\theta_D}{1.45} \exp\left(-1.04 \frac{(1+\lambda)}{(\lambda - \mu^*(1+0.62\lambda))}\right)$$

and using the value of the electron-electron coulomb repulsion parameter, $\mu^*$=0.15, the value of electron-phonon interaction parameter $\lambda$, is obtained for all the carbon substitutions studied. The extracted values of $\theta_D$ and $\lambda$ are plotted as a function of carbon fraction x in Fig.7. It can be seen that $\theta_D$ decreases from 900K in pristine $MgB_2$ to 525K in $MgB_{2-x}C_x$ for x=0.5. $\lambda$ is seen to increase correspondingly from 0.8 to 0.9. The extracted values of $\lambda$ and $\theta_D$ for pristine $MgB_2$ are in agreement with earlier estimates of these quantities from the specific heat experiments[26], similar resistivity measurements[27], phonon DOS measurements[28] and calculations[29].

**DISCUSSION**

It can be seen from Fig.4, that the $T_c$ decreases with carbon fraction upto x=0.3, beyond which it shows no significant change. Plotted in Fig.8 is our $T_c$ data along with that of other results in $MgB_{2-x}C_x$ and in $Mg_{1-x}Al_xB_2$. The rate of decrease of $T_c$ with x in our study is greater than that observed by Takenobu et.al[19] and smaller than that observed by Ahn et.al[21]. The discrepancy in the variation of $T_C$ with C fraction may have its origin in the difference between the nominal starting composition and the actual composition in the different syntheses. It has been our experience that using synthesis methods, in which the stoichiometric quantities of the initial mixture are sealed in Ta tubes under 1 atmosphere of Ar pressure, phase pure samples was not as easily obtained as in the synthesis under 50 bar of Ar pressure. In the former procedure we found the presence of $MgB_4$, accompanying a weight loss of 2-3%, which could imply presence of off-stoichiometry. We believe that the method of synthesis adopted in the present study preserves the nominal composition to a better extent. The various C substitution studies also bring out the fact that amorphous C is more reactive as compared to graphite at the temperatures employed for $MgB_2$ synthesis, which could be the reason for the absence of carbon solubility reported by Paranthaman et.al.[20]

From Fig.4 and Fig.8, it can be seen that the rate of decrease of $T_c$ with the carbon fraction, x is smaller than that observed in the case of Al. It is clear that with C substitution one electron will be added to the conduction band for every B substituted in $MgB_2$, very similar to the situation in $Mg_{1-x}Al_xB_2$. It is therefore meaningful to compare changes in $T_C$ vis-vis the cell volume in the two systems as a function of the substituting fraction x. The 'a' lattice parameter decreases in the $MgB_{2-x}C_x$ for x varying from 0.0 to 0.3 and the corresponding decrease in the lattice volume is found to be $0.65A^3$. In the case of $Mg_{1-x}Al_xB_2$ the decrease of lattice parameter occurs both along the a-axis and the c-axis[8], resulting in a decrease in the lattice volume by ~1.5 $A^3$, for a similar change in concentration. External pressure is seen to decrease $T_c$ in $MgB_2$, which in conjunction with experiments of compressibility imply a rate of decrease of $T_c$[16] with volume to be $1K/0.178A^3$. If it were lattice compression alone contributing to the decrease in $T_C$ in the C and Al substitution studies, the expected decrease in $T_c$ would be ~3.6K and ~8.6K respectively, for a fractional substitution of x upto 0.5. But from the experiments it is evident that $T_c$ decreases by 12K in the carbon substitution study and by 24K in the case of Al substitution. This difference between the observed decrease in $T_c$ and that expected from compression of the lattice can arise because of the increase in the $N(E_F)$ due to the electron donating nature of the Al and C substituents. A direct correlation between the observed decrease in $T_c$ and the increase in electron density of states at $E_F$, along the Γ-A direction has been recently pointed out by theoretical calculations for $Mg_{1-x}Al_XB_2$[18]. In contrast, in a recent study on explaining the pressure dependence of $T_C$[17], it was shown that the rate of depression of $T_c$ with pressure has two contributions, (i) due to a decrease in the electron DOS at $N(E_F)$ and (ii) an increase in average phonon frequencies $<\omega^2>$, due to stiffening of the lattice. In the context of our study in $MgB_{2-x}C_x$, it is clear that the observed decrease in $T_c$ is much larger than that expected due to volume change alone. The large decrease is seen to scale with the decrease in $\theta_D$. This may point out to the fact that $\theta_D$ may be the controlling factor in determining the $T_c$ in this system. It should be pointed out that the observed decrease in $\theta_D$, is in contrast to the calculations of Mehl et. al[6], in which an increase in the stiffness of the boron layers was predicted on account of C substitution. Independent estimates of $\theta_D$ from specific heat measurements in the carbon substituted samples will certainly be useful in clarifying this point.

The estimated $\lambda$ from the calculation is seen to increase from 0.8 to 0.9 with C substitution in our investigations (cf. Fig. 7). Using the Hopfield approximation λ is given by

$$\lambda = \frac{I * N(E_F)}{M(\langle \omega^2 \rangle)}$$

Where I is an atom related parameter, $N(E_F)$ the density of states, M the mass of the atom and $<\omega^2>$ is the average phonon frequency that couples with the electrons. Assuming that the variation of ω is similar to that of $\theta_D$, across the C series under study and M is a weighted average of the mass of C and B, to account for the observed variation in λ, it can be seen that the $N(E_F)$ should decrease by a factor of 2.6. Band structure calculations show that electron DOS increases with C/Al substitution[7,18]. Whereas the hole DOS shows a small decrease for small electron doping, and for electron doping >0.25 electrons per formula unit, the hole DOS drops drastically. Since our λ variation (cf. Fig.7), suggests a large decrease in the $N(E_F)$ with carbon substitution, it appears that it is indeed the hole DOS that is responsible for determining λ and superconductivity in this system, as as been pointed out earlier[4,5,6].

**SUMMARY AND CONCLUSIONS**

In summary we have investigated $MgB_{2-x}C_x$ for x=0.0 to x=0.50 and have shown that the carbon solubility is larger than that reported earlier[19]. This is a direct consequence of the method of preparation, in which there are no evaporation losses of Mg on account of the high pressure Ar atmosphere. The a-lattice parameter decreases with C substitution while the c-lattice parameter shows a small increase. There is an overall decrease in the lattice volume by $0.65A^{o3}$, smaller than that observed in Al for a similar composition change. The superconducting transition temperatures decrease with carbon substitution at a rate that is smaller than that seen in Al substitution. Analysis of the temperature dependence of resistivity in the normal state is well described by the Bloch-Gruneisen formula. From the above fits the Debye temperature $\theta_{D,}$ has been extracted. $\theta_D$ is seen to decrease from 900K in pristine $MgB_2$ to 525K in the x=0.5 sample. Using this $\theta_D$, and the observed value of $T_c$, the electron phonon interaction parameter, λ, has been extracted. The magnitude of $\theta_D$ and λ are in general agreement with reported values in this system. The observed increase in λ taken in conjunction with the expected decrease in $<\omega^2>$ consequent to a decrease in $\theta_D$, implies a decrease in $N(E_F)$, which taken in conjunction with band calculations implies a decrease in hole DOS with C substitution.

**FIGURE CAPTIONS**

Fig. 1a. XRD patterns in the 2θ range $10^o$ - $70^o$ for various C substitutions.

Fig. 1b The (100) peak as a function of C fraction, the intensity axis has been zero shifted for clarity.

Fig. 2(a), (b) and (c). The a- and c- lattice parameters and the lattice volume extracted from the XRD data as a function of C fraction.

Fig.3. The diamagnetic susceptibility in the 4K to 50K range for the various C fractions.

Fig.4 The variation of $T_C$ as a function of C fraction as extracted from ac susceptibility and resistivity.

Fig.5 Variation of resistivity in the 4K to 300K range for all C fractions studied. Inset shows a closeup at $T_C$. The Bloch-Gruneisen formula fits in the 40K and 300K range for all samples are shown as solid lines.

Fig.6. Variation in the ρ(300K)/ρ(40K) with $T_C/T_m$ in the C substituted samples shown as full circles. The open circles are the corresponding values for existing data in $MgB_2$ from Ref 25.

Fig.7. Variation of $\theta_D$ from the R(T),Bloch-Gruneisen formula fits versus C fraction. The variation of λ using $\theta_D$ in the McMillan equation for the various C fractions, the lines are a guide to the eye.

Fig.8 Variation of $T_C$ in our study shown as full circles shown along with that obtained for Al substitution[9].

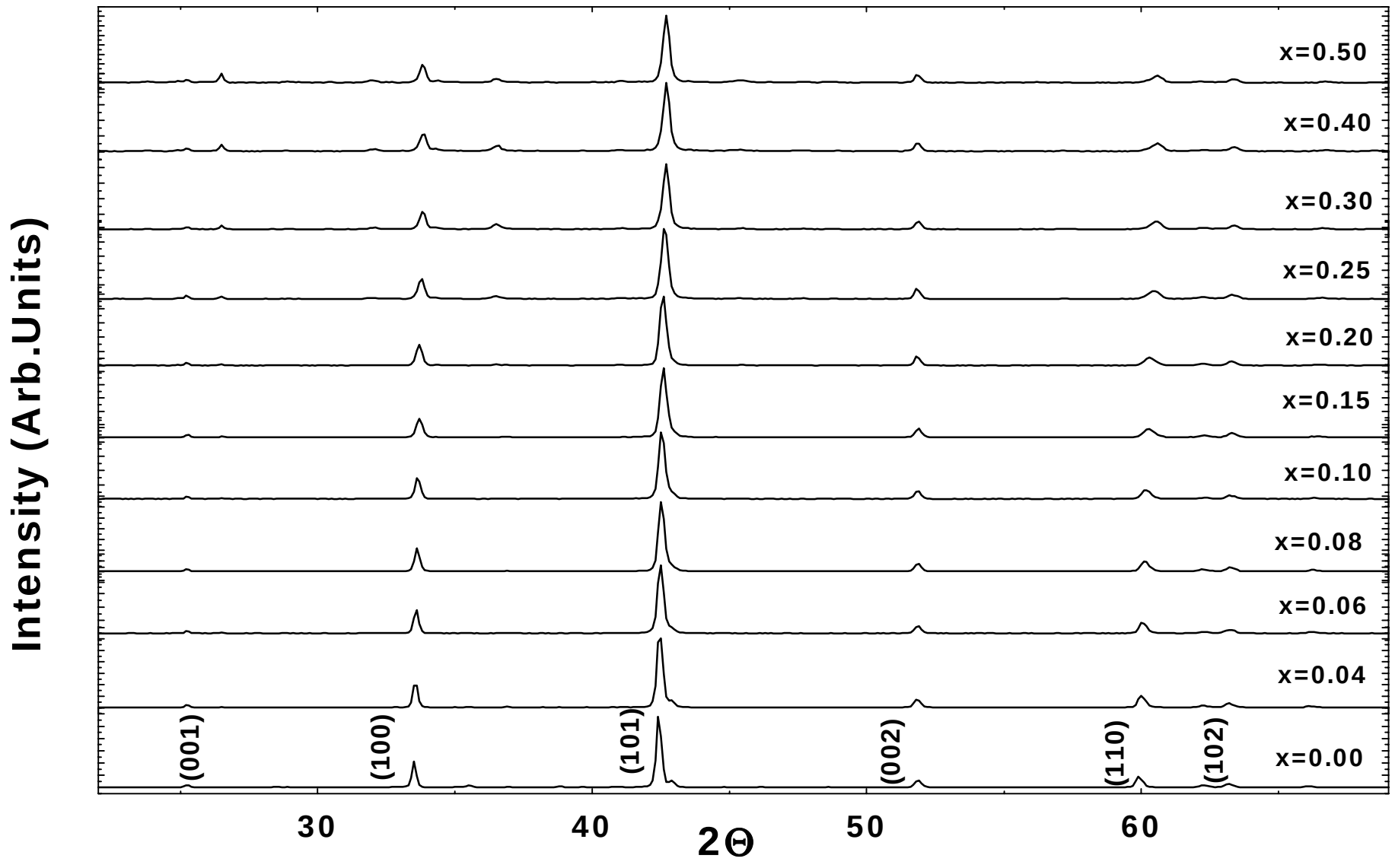

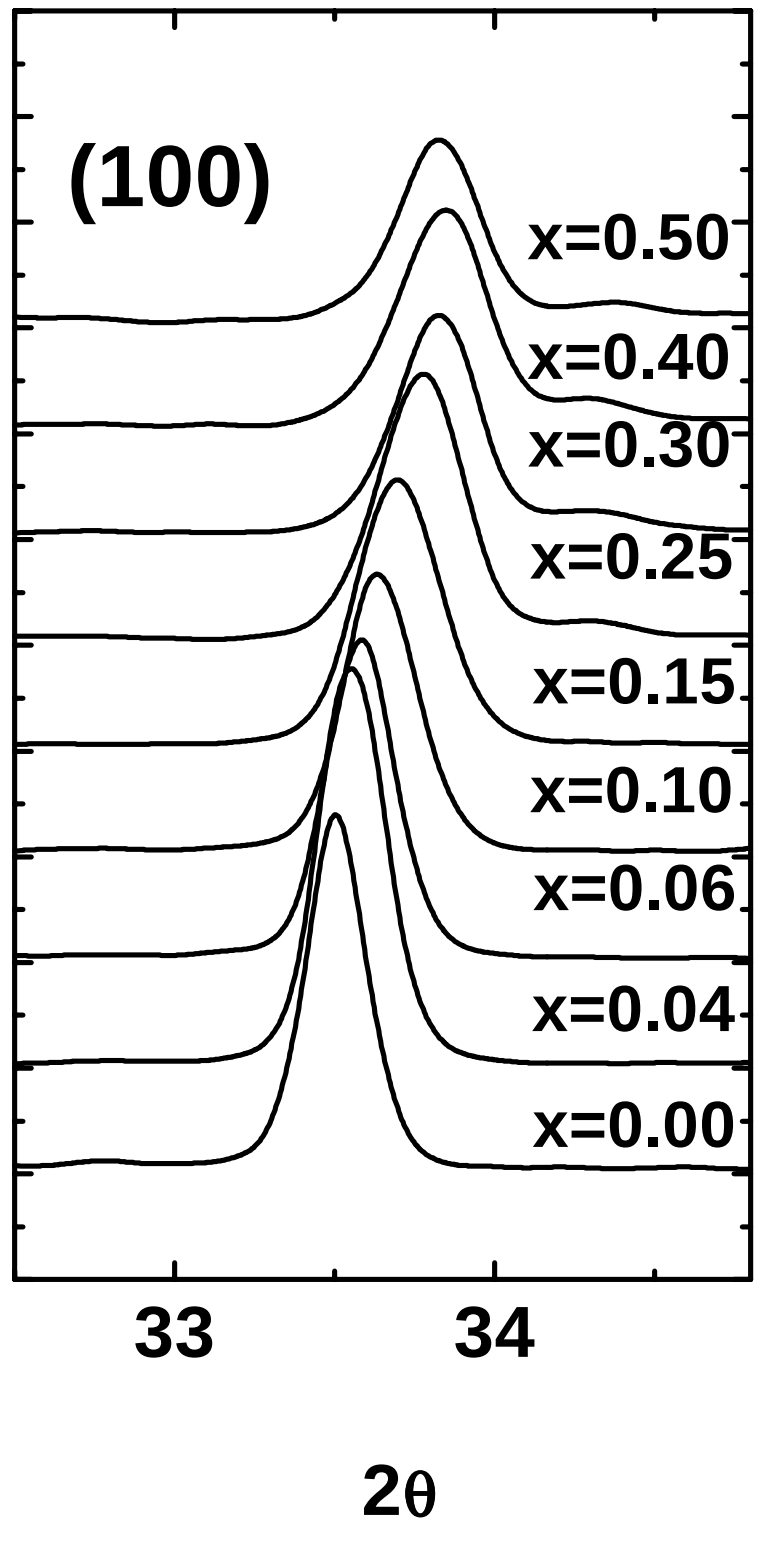

Fig. 1a. XRD patterns in the 2θ range $10^o$ - $70^o$ for various C substitutions.
Fig. 1b The (100) peak as a function of C fraction, the intensity axis has been zero shifted for clarity.

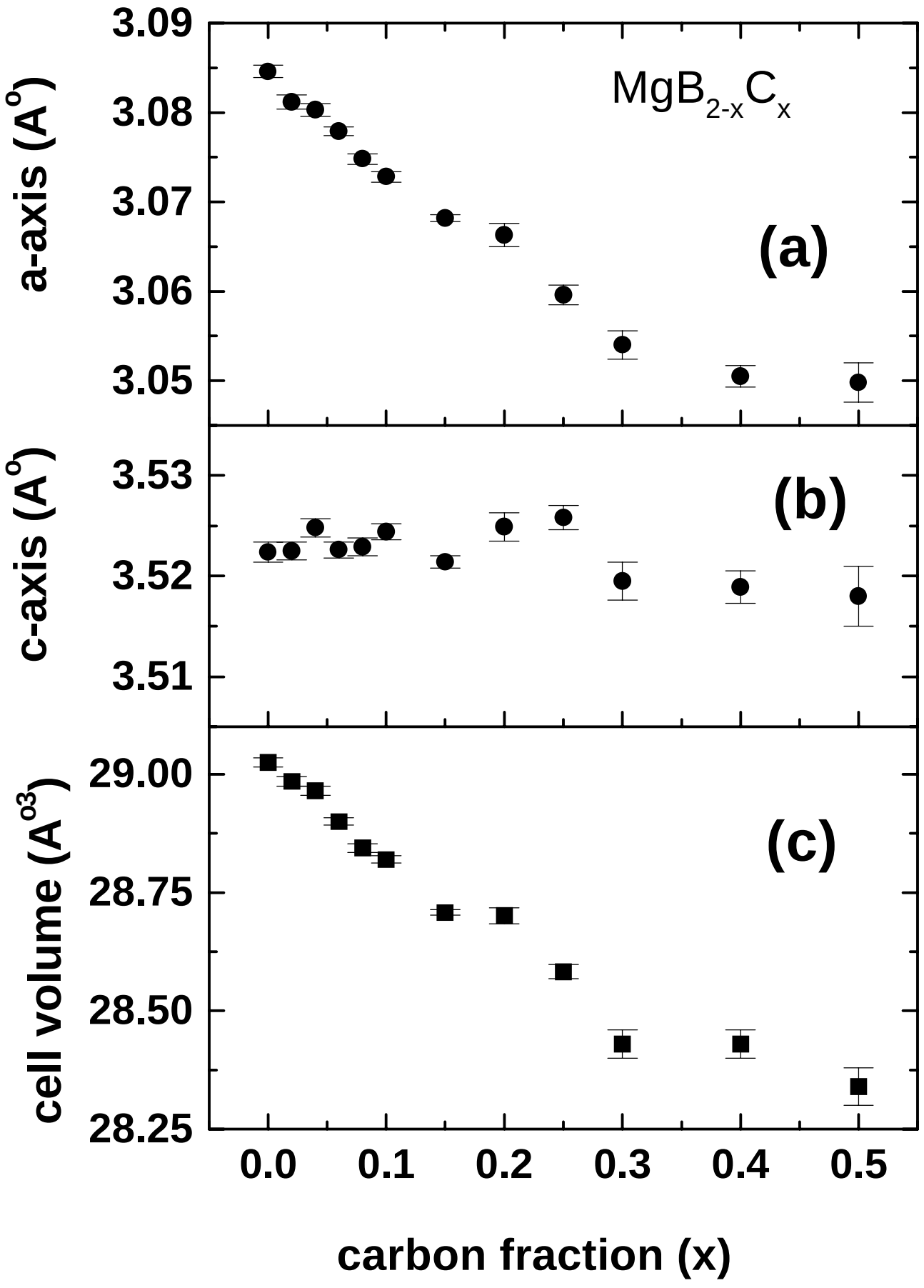

Fig. 2(a), (b) and (c). The a- and c-lattice parameters and the lattice volume extracted from the XRD data as a function of C fraction.

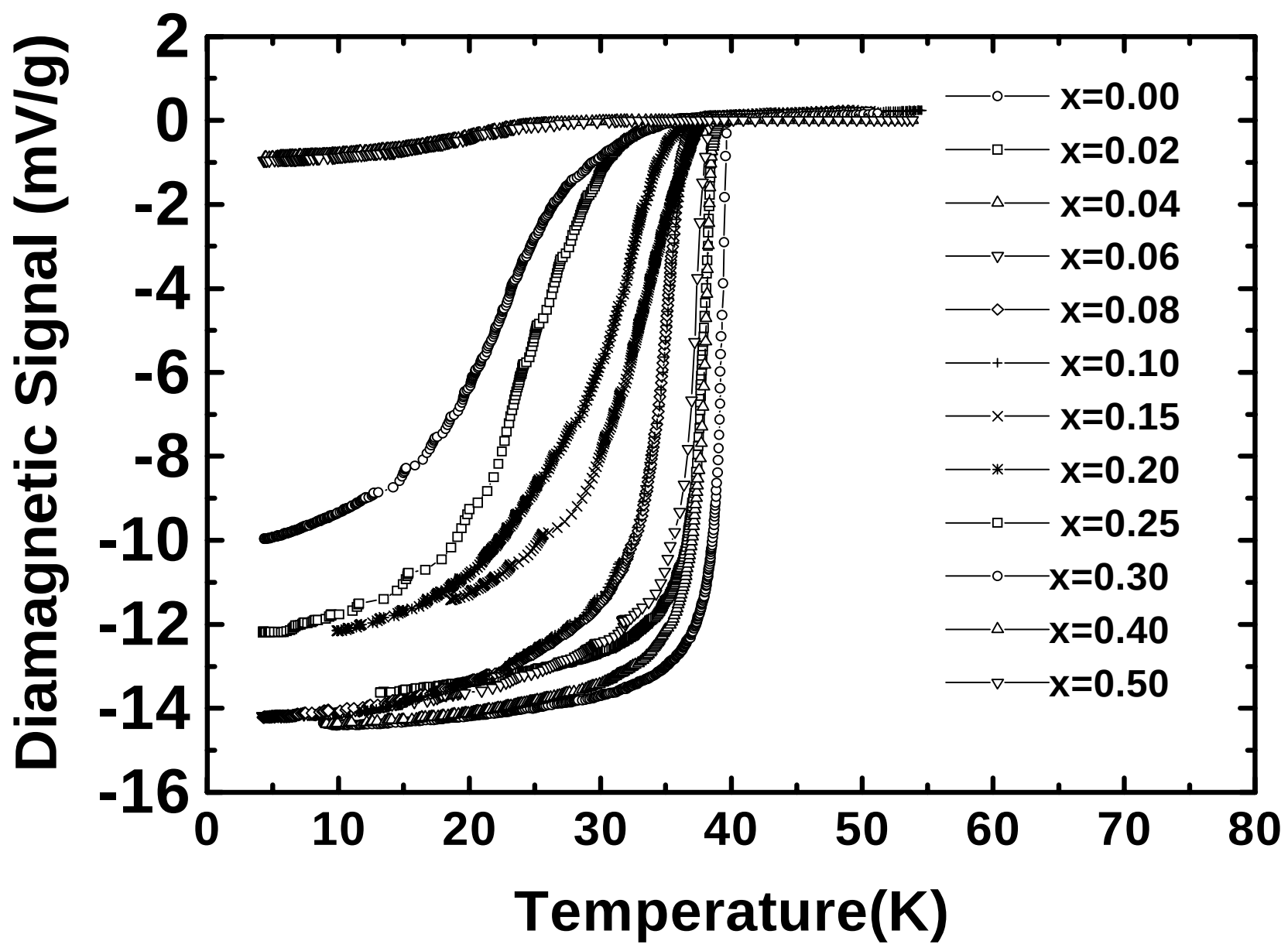

Fig.3. The diamagnetic susceptibility in the 4K to 50K range for the various C fractions.

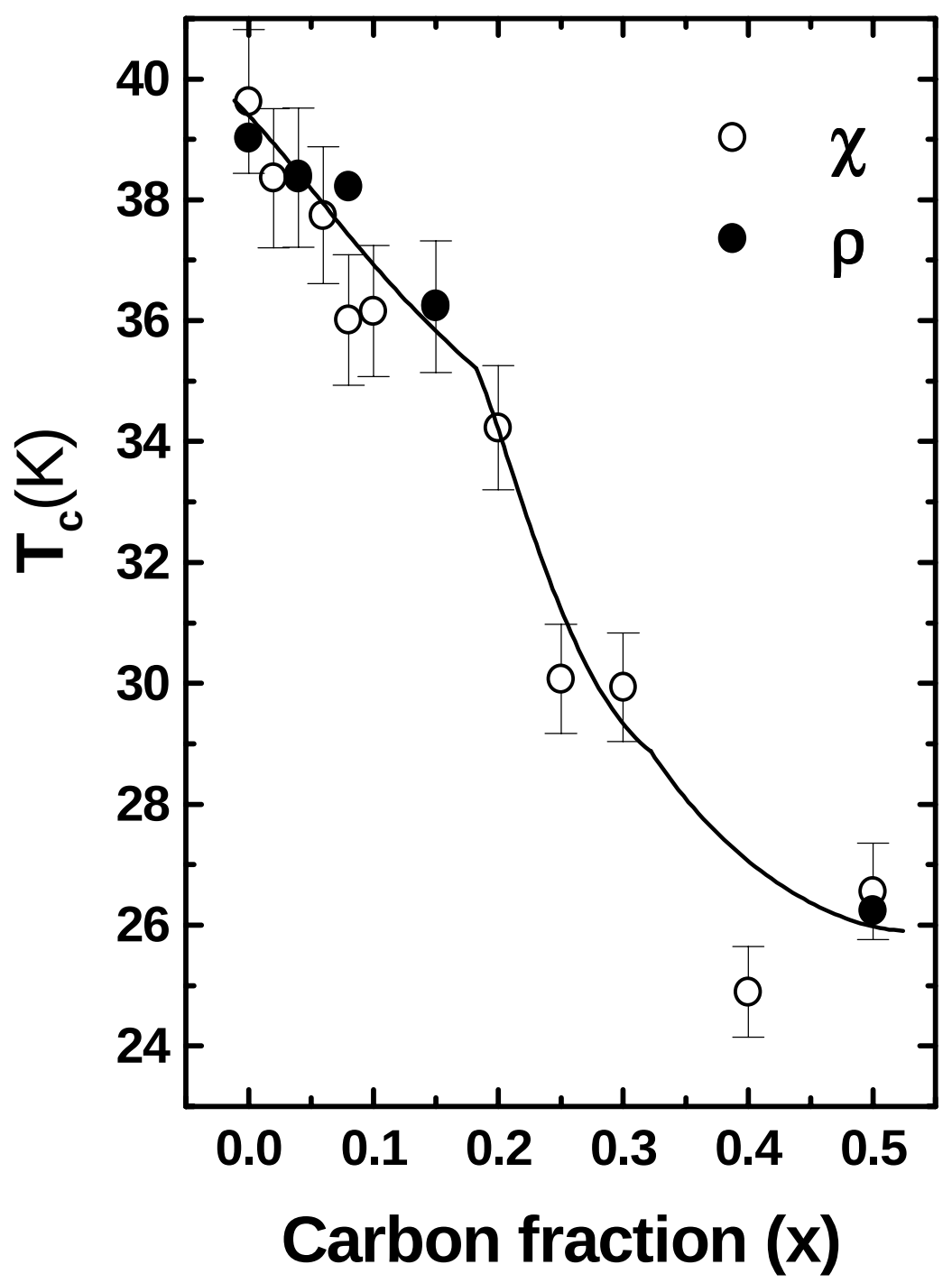

Fig.4 The variation of $T_C$ as a function of C fraction as extracted from ac susceptibility and resistivity.

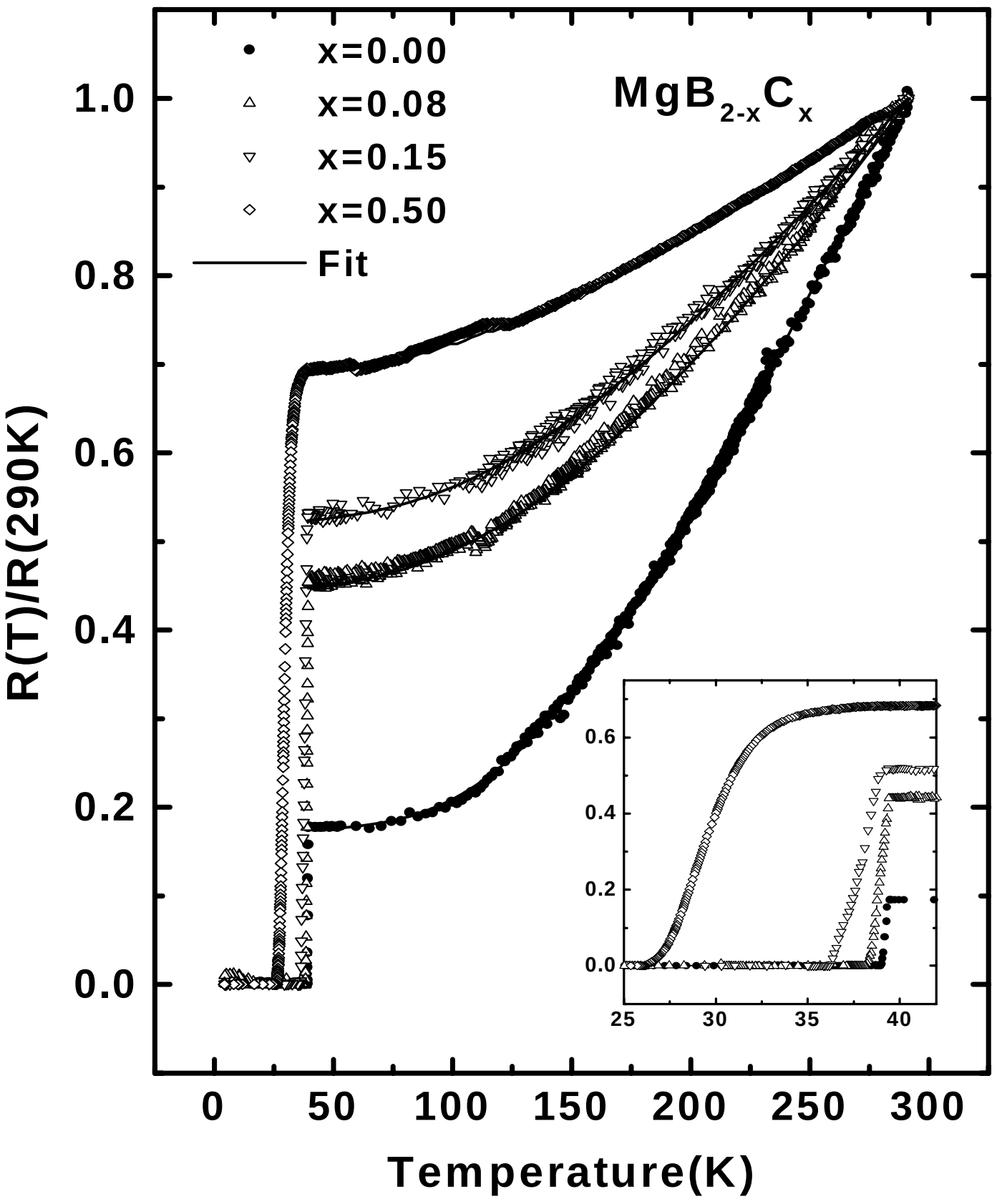

Fig.5 Variation of resistivity in the 4K to 300K range for all C fractions studied. Inset shows a closeup at $T_C$. The Bloch-Gruneisen formula fits in the 40K and 300K range for all samples are shown as solid lines.

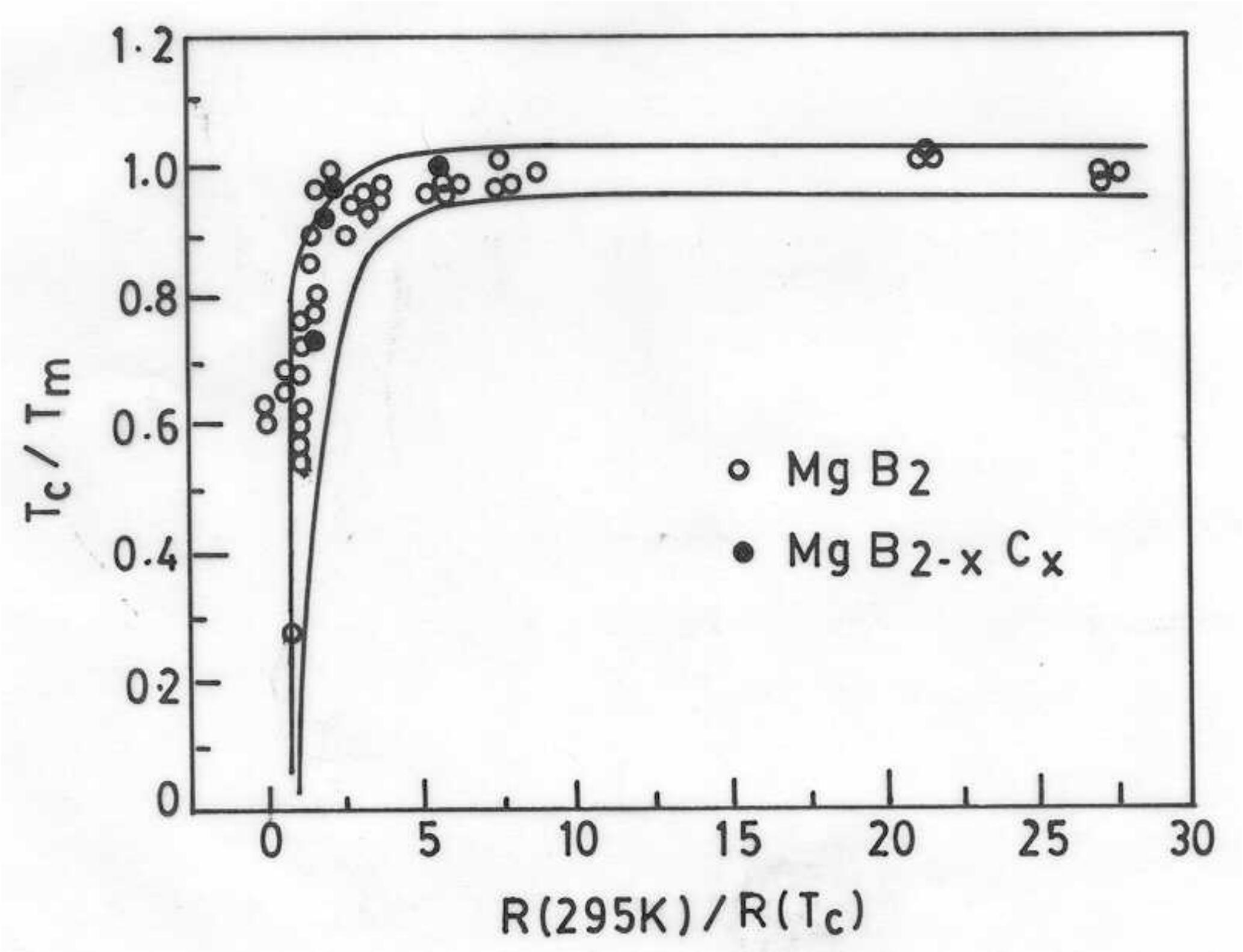

Fig.6. Variation in the R(300K)/R(40K) with $T_C/T_m$ in the C substituted samples shown as full circles. The open circles are the corresponding values for existing data in $MgB_2$ from Ref 25.

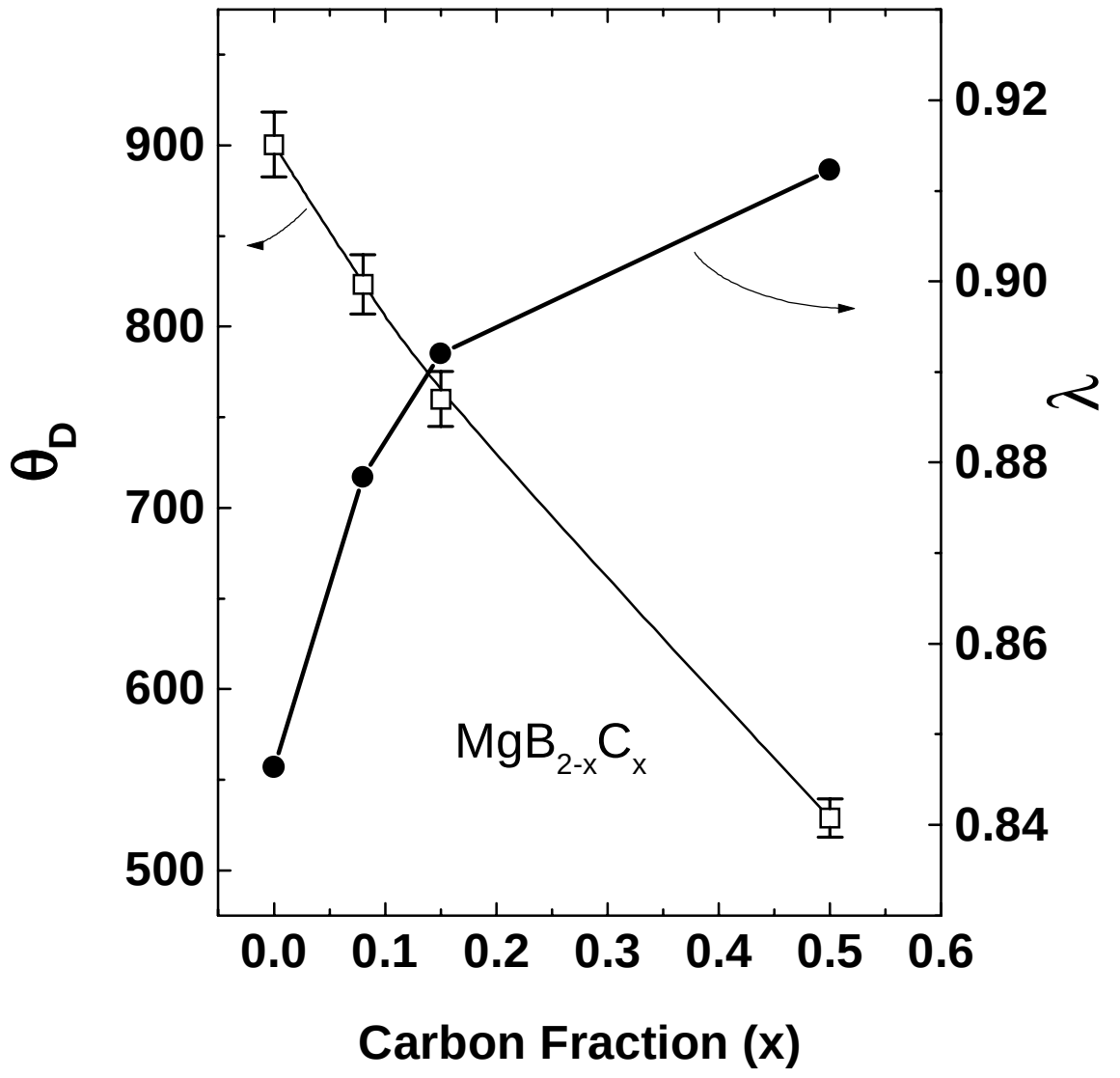

Fig.7. Variation of $\theta_D$ from the R(T),Bloch-Gruneisen formula fits versus C fraction. The variation of $\lambda$ using $\theta_D$ in the McMillan equation for the various C fractions, the lines are a guide to the eye.

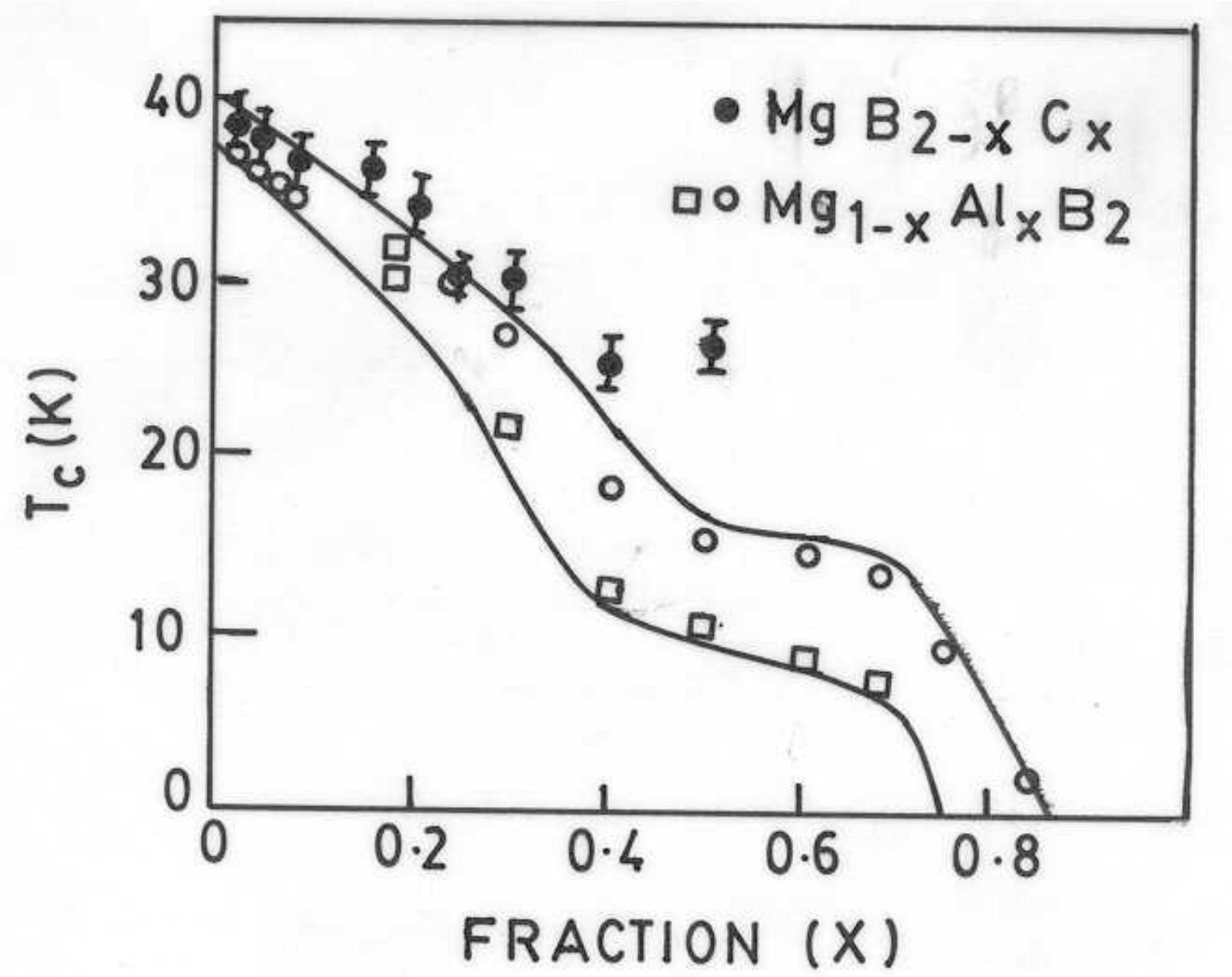

Fig.8 Variation of $T_C$ in our study shown as full circles shown along with that obtained for Al substitution[9].